\documentclass[prd,tightenlines,showpacs]{revtex4}
\usepackage{epsfig}
\usepackage{graphics}
\usepackage{latexsym}
\usepackage{amsmath}
\usepackage{amssymb}
\usepackage{rotating}
\usepackage{subfigure}
\usepackage{multirow}

\newcommand{\ar}{\arrowvert}

\newcommand{\be}{\begin{equation}}
\newcommand{\ee}{\end{equation}}
\newcommand{\ba}{\begin{eqnarray}}
\newcommand{\ea}{\end{eqnarray}}
\renewcommand{\slash}{ \not}

\begin{document}

\title{Unified Hamiltonian model for mesons and baryons}

\author{W. Xie$^{a}$}
\author{P. Wang$^{ab}$}

\affiliation{$^a$Institute of High Energy Physics, CAS, P. O. Box
918(4), Beijing 100049, China}

\affiliation{$^b$Theoretical Physics Center for Science Facilities,
CAS, Beijing 100049, China}

\begin{abstract}

A new Hamiltonian model is introduced to study the spectrum of light hadrons. It combines relativistic field theory with elements of the  constituent quark model. In addition to the standard linear confining and pseudoscalar meson exchange interactions with predetermined parameters,  an additional interaction
with different covariant spin structures is examined.
Using a large scale Monte Carlo variational procedure, the resulting model Hamiltonian provides a very good, unified description of the light quark baryon (both octet and decuplet) and  meson spectra.
\end{abstract}


\maketitle

\section{Introduction}

Although QCD is widely accepted as the fundamental theory for the strong interaction, it is extremely challenging to calculate the observed hadron spectrum directly from the QCD Lagrangian.
Due to non-Abelian and nonperturbative aspects, one  method  is  lattice QCD which provides reasonably well described ground states \cite{Allton:1998gi,Aoki:1999yr,Leinweber:1999ig}
and some success for excited states \cite{Sasaki:2001nf, Edwards:2011jj, Melnitchouk:2002eg}.
However for light hadrons such as the pion and even the proton, accurate predictions are still not possible.
Consequently there are many phenomenological models and effective theories, such as QCD sum rules
\cite{Shifman:1978bx,Shifman:1978by,Ioffe:1981kw,Dai:2003yg,Wang:2006ida}, NRQCD \cite{Brambilla:1999xf, Kniehl:2002br,Pineda:2001ra},
chiral perturbation theory \cite{Jenkins:1991ts,GuoFK2,Wang2012,Wang:2013kva,GuoFK} along with
potential models \cite {Godfrey:1998pd,Eichten:2007qx,DGG}.
Typical potential models utilize a Cornell type  interaction having  linear confinement supplemented with the usual Coulomb potential governing  short-distance behavior. A representative example is the constituent model detailed in Ref. \cite{Godfrey:1985xj} which,
with an  additional spin dependent interaction, obtained a good description of the light meson spectrum.
This model has also been  applied to baryons   \cite{Capstick:1986bm} with similar success but required different model parameters.
To describe the  interaction between color singlet objects potential models have been extended by  including a meson-exchange interaction, such as
the chiral $SU(3)$ quark potential model
 \cite{Glozman:1995fu,Zhang:1994pp,Zhang:1997ny,DaiLR}
that gave a good description of  the baryon interaction.

A  more theoretical and less phenomenological approach is the Coulomb gauge model which has been successfully applied to
 mesons \cite{LlanesEstrada:1999uh,LlanesEstrada:2001kr,LlanesEstrada:2004},
glueballs \cite{Szczepaniak:1995cw, Cotanch:1998ph}, hybrids \cite{Cotanch:2001mc,LlanesEstrada:2000hj,General:2006ed} and
tetraquark states \cite{General:2007bk,Wang:2008mw,Xie}. The predicted results are consistent
with both lattice simulations and experimental data. Different from the above potential models, the Coulomb gauge  approach entails relativistic field theory and is formulated in the same mathematical framework as the exact QCD Hamiltonian in the Coulomb gauge.
Further, it contains no free
model parameters as it only utilizes the known current quark masses and two dynamical constants, the string tension $\sigma$ and the QCD coupling constant $\alpha_s$,
that are predetermined from the literature. While this model provides a reasonable hadron description it does not simultaneously reproduce both the meson and baryon spectrums with the same overall accuracy as the multi-parameter purely meson or baryon  models mentioned above.

The purpose of the current work is to provide  a unified model that can accurately reproduce both meson and baryon masses with the same set of model parameters which to date has not been achieved.  The motivation is to develope a robust framework for reliably predicting and understanding more exotic systems such as light and heavy tetraquark states which are of intense interest.
Building on the attractive theoretical features of the Coulomb gauge model and phenomenological successes of  constituent quark models, a unified Hamiltonian approach has been developed
that combines relativity, field theory and elements of the constituent quark model with, most significantly, a single set of parameters
that can simultaneosuly describe both meson and baryon masses.

This paper is organized into six sections. In section \ref{sec:model} the unified Hamiltonion is described
and then the meson and baryon model wavefunctions are detailed
 in sections  \ref{sec:mesons} and  \ref{sec:baryons}, respectively.
Section \ref{sec:results} presents numerical results and highlights the accurate hadron descriptions. Finally,  conclusions are summarized in section \ref{sec:conclusion}.
\section{Model Hamiltonian}\label{sec:model}

The model Hamiltonian is
\begin{eqnarray}
H_{t} &=& H_{kine} + H_{I0} + H_{I1} + H_{ch},
\end{eqnarray}
\label{model}
where
$H_{kine}$ is  the relativistic kinetic energy
\begin{equation}
H_{kine}= \int d{\bf x} \Psi^{\dag} ({\bf x}) (-i\alpha\cdot\nabla + \beta m) \Psi ({\bf x}),
\end{equation}
and $H_{I0}$, $H_{I1}$, and  $H_{ch}$ are the interactions detailed below.
Similar to the Coulomb gauge
model \cite{LlanesEstrada:1999uh,LlanesEstrada:2001kr},
$H_{I0}$ is the confining interaction
\begin{equation}
H_{I0}= -\frac{1}{2} \int d{\bf x} d{\bf y} [\bar{ \Psi}({\bf x})\gamma^0 T^{a}\Psi({\bf x})] V_0(\ar {\bf x}-{\bf y} \ar ) [\bar{ \Psi}({\bf y})\gamma^0 T^{a}\Psi({\bf y})],
\end{equation}
where $T^a=\frac{\lambda^a}{2}$ are the color $SU(3)$ group generators and $V_0(\ar {\bf x}-{\bf y} \ar )$ is a Cornell type potential
\be
 V_0(\ar {\bf x}-{\bf y} \ar)=\frac{C}{(2\pi)^3}-\frac{\alpha_s}{\ar {\bf x}-{\bf y} \ar}+\sigma \ar {\bf x}-{\bf y} \ar.
\ee
Following constituent quark models a constant energy $C$ is introduced and $\sigma$, $\alpha_s$ are  the same as in the Coulomb gauge  model.
This is a ``charge-charge" color interaction. Performing a Fourier transformation, the potential in momentum space is
\begin{equation}
V_0(|{\bf q}|)= C \delta^3 ({\bf q}) - \frac{4\pi\alpha_s}{q^2} - \frac{8\pi\sigma}{q^4} + \delta^3 ({\bf q})\int d {\bf q}'\frac{8\pi\sigma}{q'^4}.
\end{equation}
The last term is to satisfy the condition that at $r=0$, the confining potential is zero \cite{Gross}. It is also important to
 deal with the divergence of the integral with linear confining potential at zero momentum transfer.

The interaction between two colored objects can have other forms, for example, ``current-current" interaction,
``spin-spin" interaction, etc. In particular to account for hadron spin splittings a hyperfine type interaction $H_{I1}$ is included with structure
\begin{equation}
H_{I1}=-\frac{1}{2} \int d{\bf x} d{\bf y} [\bar{ \Psi}({\bf x})\Gamma T^{a}\Psi({\bf x})] V_1(\ar {\bf x}-{\bf y} \ar ) [\bar{ \Psi}({\bf y})\Gamma T^{a}\Psi({\bf y})].
\end{equation}
The $\Gamma$ matrix can be $1,\vec{\gamma},\gamma_5,\vec{\gamma}\gamma_0,\gamma_5\gamma_0,\vec{\gamma}\gamma_5$.
The potential $V_1(\ar {\bf x}-{\bf y} \ar )$ is taken to be similar to $V_0$ with linear and Coulomb terms
\ba
 V_1(\ar {\bf x}-{\bf y} \ar)=-\frac{\alpha_1}{\ar {\bf x}-{\bf y} \ar}+\sigma_1 \ar {\bf x}-{\bf y} \ar.\\
\ea
This interaction will be used to
 reproduce the $\pi$$\rho$ splitting which is large due to the small $\pi$ mass governed by chiral symmetry as documented in Ref. \cite{LlanesEstrada:1999uh} which uses a Random Phase Approximation diagonalization to obtain a light chiral pion. Here a light pion mass is obtained entirely via spin-splitting similar to the constituent treatment of Ref.  \cite{Godfrey:1985xj}.

The above interaction is between two colored objects.
To describe interacting color singlet hadrons a pseudoscalar meson exchange interaction $H_{ch}$
is also included using the quark-meson  Lagrangian
\be
{\cal L}_{ch}=-g_{ch}\overline{\psi}
(i\gamma_5\sum^8_{a=1}\lambda_a\pi_a)\psi.
\ee
Here $\lambda_a$ are the  $SU_f(3)$ generators and $\pi_a$ are the pseudoscalar meson fields.
The coupling constant $g_{ch}$ is determined from the $NN\pi$ interaction \cite{Zhang:1997ny}
\be
\frac{g^2_{ch}}{4\pi}=\frac{9}{25}\frac{m^2_u}{m^2_N} \frac{g^2_{NN\pi}}{4\pi},
\ee
where $\frac{g^2_{NN\pi}}{4\pi} = 13.67$ \cite{Bugg:2004cm}.
The constituent quark mass $m_u$ is chosen to be 220 MeV \cite{Godfrey:1985xj,Capstick:1986bm}.
The Goldstone field is
  \begin{equation}
\sum^8_{a=1}\lambda_a\pi_a=
\left(
  \begin{array}{ccc}
\pi^0+\frac{1}{\sqrt{3}}\eta & \sqrt 2\pi^+ &  \sqrt 2 K^+ \\
 \sqrt 2 \pi^- & -\pi^0+\frac{1}{\sqrt{3}}\eta &  \sqrt 2 K^0 \\
 \sqrt 2 K^- &  \sqrt 2 \bar{K}^0 & -\frac{2}{\sqrt 3}\eta \\
  \end{array}
\right).
\end{equation}
From the quark-meson interaction the one-meson exchange potential can be extracted.
For example, the one-pion exchange potential between two color singlets is
\be
H_{ch}^{\pi}=\frac{1}{2} \int d{\bf x} d{\bf y} [\bar{ \Psi}({\bf x})\gamma^5 \Psi({\bf x})] V_{ch}^{\pi}(\ar {\bf x}-{\bf y} \ar ) [\bar{ \Psi}({\bf y})\gamma^5\Psi({\bf y})],
\ee
where $V_{ch}^{\pi}(\ar {\bf x}-{\bf y} \ar )$ is the Fourier transformation of $\hat V_{ch}^{\pi}(\mathbf q)$
\be
\hat V_{ch}^{\pi}(\mathbf q)=\frac{ g^2_{ch}}{{\bf q}^2+m^2_{\pi}}.
\ee
The pseudoscalar meson mass in the meson exchange potential is chosen to be the experimental value.

In the above equations, the quark field operators can be expanded
\begin{eqnarray}
\label{colorfields1}
 \Psi(\bf{x}) &=&\int \!\! \frac{d
    \bf{k}}{(2\pi)^3} [{u}_{\lambda}
({\bf k}) b_{\lambda \cal C}({\bf k})  +
{v}_{\lambda} (-{\bf k})
    d^\dag_{\lambda {\cal C}}(-{\bf k})]  e^{i {\bf k} \cdot \bf{x}} {\boldsymbol{\epsilon}}_{\cal C},  \\
 \bar \Psi(\bf{x}) &=&\int \!\! \frac{d
    \bf{k}}{(2\pi)^3} [{\bar u}_{\lambda}
({\bf k}) b^{\dagger}_{\lambda \cal C}({\bf k})  +
{\bar v}_{\lambda} (-{\bf k})
    d_{\lambda {\cal C}}(-{\bf k})]  e^{i {\bf k} \cdot \bf{x}} {\boldsymbol{\bar \epsilon}}_{\cal C},
\end{eqnarray}
where the Dirac spinors are
\begin{eqnarray}
 u_{\lambda}({\bf k})&=&\frac{1}{\sqrt 2}
 \left(
  \begin{array}{c}
    \sqrt{\frac{\omega+m}{\omega}}\chi_{\lambda}\\
     \sqrt{\frac{\omega-m}{\omega}}\boldsymbol{\sigma}\cdot\hat{\bf k}\chi_{\lambda}
  \end{array}
  \right),\\
 v_{\lambda}({\bf k})&=&\frac{1}{\sqrt 2}
 \left(
  \begin{array}{c}
    \sqrt{\frac{\omega-m}{\omega}} \boldsymbol{\sigma}\cdot\hat{\bf k} \}\\
     \sqrt{\frac{\omega+m}{\omega}}\chi_{\lambda}
  \end{array}
  \right),
\end{eqnarray}
and  $\omega=\sqrt{m^2+{\bf k}^2}$. The spinors $\chi_{\lambda}$ are,  fermions,
 $
\chi_{+} =\left(
  \begin{array}{c}
    1\\
    0
  \end{array}
  \right)
$,
 $
\chi_{-} =\left(
  \begin{array}{c}
    0\\
    1
  \end{array}
  \right)
$
and anti-fermions,
 $
\chi_{+} =\left(
  \begin{array}{c}
    0\\
    1
  \end{array}
  \right)
$,
 $
\chi_{-} =\left(
  \begin{array}{c}
    1\\
    0
  \end{array}
  \right)
$.

\section{Mesons}\label{sec:mesons}

 In the center of momentum  the meson state $|q \bar {q} \rangle$ is given by
\begin{equation}
\arrowvert \Psi^{J^{PC}} \rangle =\sum_{\mathcal C_1 \mathcal C_3 \lambda_1 \lambda_3}\int
\frac{d\mathbf{k}}{(2\pi)^3}
\Psi ^{J^{PC}}_{\mathcal C_1 \mathcal C_{3} \lambda_1 \lambda_3}(\mathbf{k}) b^{\dagger}_{\mathcal C_1 \lambda_1}(\mathbf{k})
d^{\dagger}_{\mathcal C_3 \lambda_3}(-\mathbf{k})
\arrowvert
0 \rangle,
\end{equation}
with  convention  1 for a quark having momentum $\mathbf{k}$ and  3 for an anti-quark having momentum $\mathbf{-k}$. Color and spin are represented by  $\mathcal C$ and $\lambda$,  respectively.  The wavefunction $\Psi ^{J^{PC}}_{\mathcal C_1 \mathcal C_{3} \lambda_1 \lambda_3}(\mathbf{k})$ has form
\begin{equation}
	\Psi ^{J^{PC}}_{\mathcal C_1 \mathcal C_{3} \lambda_1 \lambda_3}(\mathbf{k})=
\delta_{\mathcal C_1 \mathcal C_3} f(k)
\sum_{m_L m_S}
 \langle \frac{1}{2}  \frac{1}{2} \lambda_1 \lambda_3\arrowvert S m_S \rangle
 \langle L S m_L m_S \arrowvert J m_J \rangle
(-1)^{\frac{1}{2}-\lambda_3}
Y^{m_L}_L(\mathbf{k}),
\end{equation}
with  spin and angular momentum coupling $\hat S_1+\hat S_3=\hat S$, $\hat L+\hat S=\hat J$,  $Y^{m}_L(\mathbf k)$ is the spherical harmonic function and $f(k)$ is the radial wavefunction with variational parameter $\alpha$
\ba
f(k)=k^{2L}e^{-\frac{k^2}{\alpha}}.
\ea

The meson mass is given by
\begin{eqnarray}
M&=&\frac{\langle\Psi^{J^{PC}}|H_{t}|\Psi^{J^{PC}}\rangle}{\langle\Psi^{J^{PC}}|\Psi^{J^{PC}}\rangle} \notag \\
&=&\frac{\langle\Psi^{J^{PC}}|H_{kine}|\Psi^{J^{PC}}\rangle+\langle\Psi^{J^{PC}}|H_{I0}|\Psi^{J^{PC}}\rangle +\langle\Psi^{J^{PC}}|H_{I1}|\Psi^{J^{PC}}\rangle+\langle\Psi^{J^{PC}}|H_{ch}|\Psi^{J^{PC}}\rangle}{\langle\Psi^{J^{PC}}|\Psi^{J^{PC}}\rangle} \notag\\
&=&M_{kine}+M_0+M_1+M_{ch},
\end{eqnarray}
where
\ba
M_{kine}&=&\int \frac{d\mathbf{k}}{(2\pi)^3} (\sqrt{m^2_1+{\mathbf k}^2}+\sqrt{m^2_3+{\mathbf k}^2})
 f^2(k)  \notag \\
&&\times \sum_{\lambda_1 \lambda_3 m_L m_S}
\left[\langle \frac{1}{2}  \frac{1}{2} \lambda_1 \lambda_3\arrowvert S m_S \rangle
 \langle L S m_L m_S \arrowvert J m_J \rangle \right ]^2
Y^{*m_L}_{L}(\mathbf{k})
Y^{m_L}_L(\mathbf{k}),
\ea
\ba
M_0&=&\frac{4}{3}\int \frac{d\mathbf{k}}{(2\pi)^3} \frac{d\mathbf{k}^\prime}{(2\pi)^3} {\hat V_0}(|\mathbf k-\mathbf k^\prime|) f(k)f(k^\prime) \notag\\
&&\times \sum_{\lambda_1 \lambda_3 m_L m_S}\sum_{\lambda^\prime_1 \lambda^\prime_3 m^\prime_L m^\prime_S}
\langle \frac{1}{2}  \frac{1}{2} \lambda^\prime_1 \lambda^\prime_3\arrowvert S^\prime m^\prime_S \rangle
\langle L^\prime S^\prime m^\prime_L m^\prime_S \arrowvert J^\prime m^\prime_J \rangle
 \langle \frac{1}{2}  \frac{1}{2} \lambda_1 \lambda_3\arrowvert S m_S \rangle
 \langle L S m_L m_S \arrowvert J m_J \rangle       \notag \\
&& (-1)^{\frac{1}{2}-\lambda_3+\frac{1}{2}-\lambda^\prime_3}
Y^{*m^\prime_L}_{L^\prime}(\mathbf{k^\prime})
Y^{m_L}_L(\mathbf{k})
\left[  \bar u_{\lambda_1^{\prime}} ({\bf k}^{\prime}) \gamma^0 u_{\lambda_1}({\bf k})  \right]
\left[  \bar v_{\lambda_3}({-\bf k}) \gamma^0 v_{\lambda_3^{\prime}}({-\bf k}^{\prime})  \right],
\ea
\ba
M_1&=&\frac{4}{3}\int \frac{d\mathbf{k}}{(2\pi)^3} \frac{d\mathbf{k}^\prime}{(2\pi)^3} {\hat V_1}(|\mathbf k-\mathbf k^\prime|) f(k)f(k^\prime) \notag\\
&&\times \sum_{\lambda_1 \lambda_3 m_L m_S}\sum_{\lambda^\prime_1 \lambda^\prime_3 m^\prime_L m^\prime_S}
\langle \frac{1}{2}  \frac{1}{2} \lambda^\prime_1 \lambda^\prime_3\arrowvert S^\prime m^\prime_S \rangle
\langle L^\prime S^\prime m^\prime_L m^\prime_S \arrowvert J^\prime m^\prime_J \rangle
 \langle \frac{1}{2}  \frac{1}{2} \lambda_1 \lambda_3\arrowvert S m_S \rangle
 \langle L S m_L m_S \arrowvert J m_J \rangle       \notag \\
&& (-1)^{\frac{1}{2}-\lambda_3+\frac{1}{2}-\lambda^\prime_3}
Y^{*m^\prime_L}_{L^\prime}(\mathbf{k^\prime})
Y^{m_L}_L(\mathbf{k})
\left[  \bar u_{\lambda_1^{\prime}} ({\bf k}^{\prime}) \Gamma u_{\lambda_1}({\bf k})  \right]
\left[  \bar v_{\lambda_3}({-\bf k}) \Gamma v_{\lambda_3^{\prime}}({-\bf k}^{\prime})  \right].
\ea
The contribution from the meson-exchange interaction  is
\ba
M_{ch}&=&\frac{4}{3}\int \frac{d\mathbf{k}}{(2\pi)^3} \frac{d\mathbf{k}^\prime}{(2\pi)^3} {\hat V_{ch}}(|\mathbf k-\mathbf k^\prime|) f(k)f(k^\prime) \notag\\
&&\times \sum_{\lambda_1 \lambda_3 m_L m_S}\sum_{\lambda^\prime_1 \lambda^\prime_3 m^\prime_L m^\prime_S}
\langle \frac{1}{2}  \frac{1}{2} \lambda^\prime_1 \lambda^\prime_3\arrowvert S^\prime m^\prime_S \rangle
\langle L^\prime S^\prime m^\prime_L m^\prime_S \arrowvert J^\prime m^\prime_J \rangle
 \langle \frac{1}{2}  \frac{1}{2} \lambda_1 \lambda_3\arrowvert S m_S \rangle
 \langle L S m_L m_S \arrowvert J m_J \rangle       \notag \\
&& (-1)^{\frac{1}{2}-\lambda_3+\frac{1}{2}-\lambda^\prime_3}
Y^{*m^\prime_L}_{L^\prime}(\mathbf{k^\prime})
Y^{m_L}_L(\mathbf{k})
\left[  \bar u_{\lambda_1^{\prime}} ({\bf k}^{\prime}) \gamma_5 u_{\lambda_1}({\bf k})  \right]
\left[  \bar v_{\lambda_3}({-\bf k}) \gamma_5  v_{\lambda_3^{\prime}}({-\bf k}^{\prime})  \right].
\ea

\section{Baryons}\label{sec:baryons}

 The baryon state can be constructed using quark creation operators acting on the vacuum state
\begin{equation}
\arrowvert qqq,J^P\rangle =\sum_{\mathcal C_1 \mathcal C_2\mathcal C_3}\int
\frac{d\mathbf k_1}{(2\pi)^3}   \frac{d\mathbf k_2}{(2\pi)^3}
\Psi ^{J^{P}}_{\mathcal C_1 \mathcal C_2\mathcal C_{3} \lambda_1\lambda_2 \lambda_3 f_1 f_2 f_3}(\mathbf{k}_1,\mathbf{k}_2,\mathbf{k}_3)
 b^{\dagger}_{\mathcal C_1 \lambda_1 f_1}(\mathbf{k}_1)
 b^{\dagger}_{\mathcal C_2 \lambda_2 f_2}(\mathbf{k}_2)
 b^{\dagger}_{\mathcal C_3 \lambda_3 f_3}(\mathbf{k}_3)
\arrowvert
0 \rangle,
\end{equation}
here  $\mathcal C$, $\lambda$, $f$ represent color, spin and flavor, respectively.
The  baryon wave function,
$\Psi ^{J^{P}}_{\mathcal C_1 \mathcal C_2\mathcal C_{3} \lambda_1\lambda_2 \lambda_3 f_1 f_2 f_3}(\mathbf{k}_1,\mathbf{k}_2,\mathbf{k}_3)$,  can be written as the product of momentum, flavor-spin and color wave functions
\be
\Psi ^{J^{P}}_{\mathcal C_1 \mathcal C_2\mathcal C_{3} \lambda_1\lambda_2 \lambda_3 f_1 f_2 f_3}(\mathbf{k}_1,\mathbf{k}_2,\mathbf{k}_3)
=f(\mathbf k_1,\mathbf k_2,\mathbf k_3)\times \psi_{fs}(\lambda_1,\lambda_2, \lambda_3, f_1, f_2, f_3)\times \psi_{color}(\mathcal C_1, \mathcal C_2, \mathcal C_{3}),
\ee
here $\psi_{fs}$ is the flavor-spin wave function and $\psi_{color}$ is the color wave function.  Fermi-Dirac statistics
requires the total baryon wave function  must be antisymmetric under the exchange of quarks. The baryon color state is a singlet and is antisymmetric
\be
\psi_{color}(\mathcal C_1, \mathcal C_2, \mathcal C_{3})=\varepsilon_{\mathcal C_1\mathcal C_2\mathcal C_3}.
\ee
Hence the remaining  wave function must be symmetric. Since the ground state momentum wave function $f(\mathbf k_1,\mathbf k_2,\mathbf k_3)$ is  symmetric, the flavor-spin wave function  $\psi_{fs}$ must also be  symmetric. For example the proton and $\Delta^{+}$  flavor-spin wave functions $\psi_{fs}$ are
\ba
&&\psi_{fs} (\text{proton},\frac{1}{2})=\frac{1}{3\sqrt 2} (\uparrow \downarrow \uparrow- \downarrow \uparrow \uparrow)(udu-duu)
+\frac{1}{3\sqrt 2} (\uparrow \uparrow \downarrow- \uparrow \downarrow \uparrow)(uud-udu)
+\frac{1}{3\sqrt 2} (\uparrow \uparrow \downarrow- \downarrow\uparrow  \uparrow)(uud-duu) \notag ,\\
&&\psi_{fs}(\Delta^{+},\frac{3}{2})=\frac{1}{\sqrt 3}[u(\uparrow)u(\uparrow)d(\uparrow)+u(\uparrow)d(\uparrow)u(\uparrow)
+d(\uparrow)u(\uparrow)u(\uparrow)] \notag.
\ea

To construct a completely symmetric momentum space wave function  the momentum Jacobi coordinates are
utilized
\ba
\boldsymbol{ \rho}_{12}=\frac{1}{\sqrt 2} (\mathbf k_1-\mathbf k_1), \boldsymbol{ \lambda}_{12}=\frac{1}{\sqrt 6}(\mathbf k_1+\mathbf k_2-2\mathbf k_3), \\
\boldsymbol{ \rho}_{23}=\frac{1}{\sqrt 2} (\mathbf k_2-\mathbf k_3), \boldsymbol{ \lambda}_{23}=\frac{1}{\sqrt 6}(\mathbf k_2+\mathbf k_3-2\mathbf k_1), \\
\boldsymbol{ \rho}_{31}=\frac{1}{\sqrt 2} (\mathbf k_3-\mathbf k_1), \boldsymbol{ \lambda}_{31}=\frac{1}{\sqrt 6}(\mathbf k_3+\mathbf k_1-2\mathbf k_2).
\ea
The proper symmetric variational wave function can then be written as
\be
f(\mathbf k_1,\mathbf k_2,\mathbf k_3)=e^{- \frac{ \boldsymbol{ \rho}^2_{12} }   {\alpha^2_1}  - \frac{ \boldsymbol{ \lambda}^2_{12} }   {\alpha^2_2}  } +
e^{- \frac{ \boldsymbol{ \rho}^2_{23} }   {\alpha^2_1}  - \frac{ \boldsymbol{ \lambda}^2_{23} }   {\alpha^2_2}  }
+e^{- \frac{ \boldsymbol{ \rho}^2_{31} }   {\alpha^2_1}  - \frac{ \boldsymbol{ \lambda}^2_{31} }   {\alpha^2_2}  },
\ee
where $\alpha_1$ and $\alpha_2$ are determined by the variational method.

\begin{figure}[t]
\includegraphics[width=12cm]{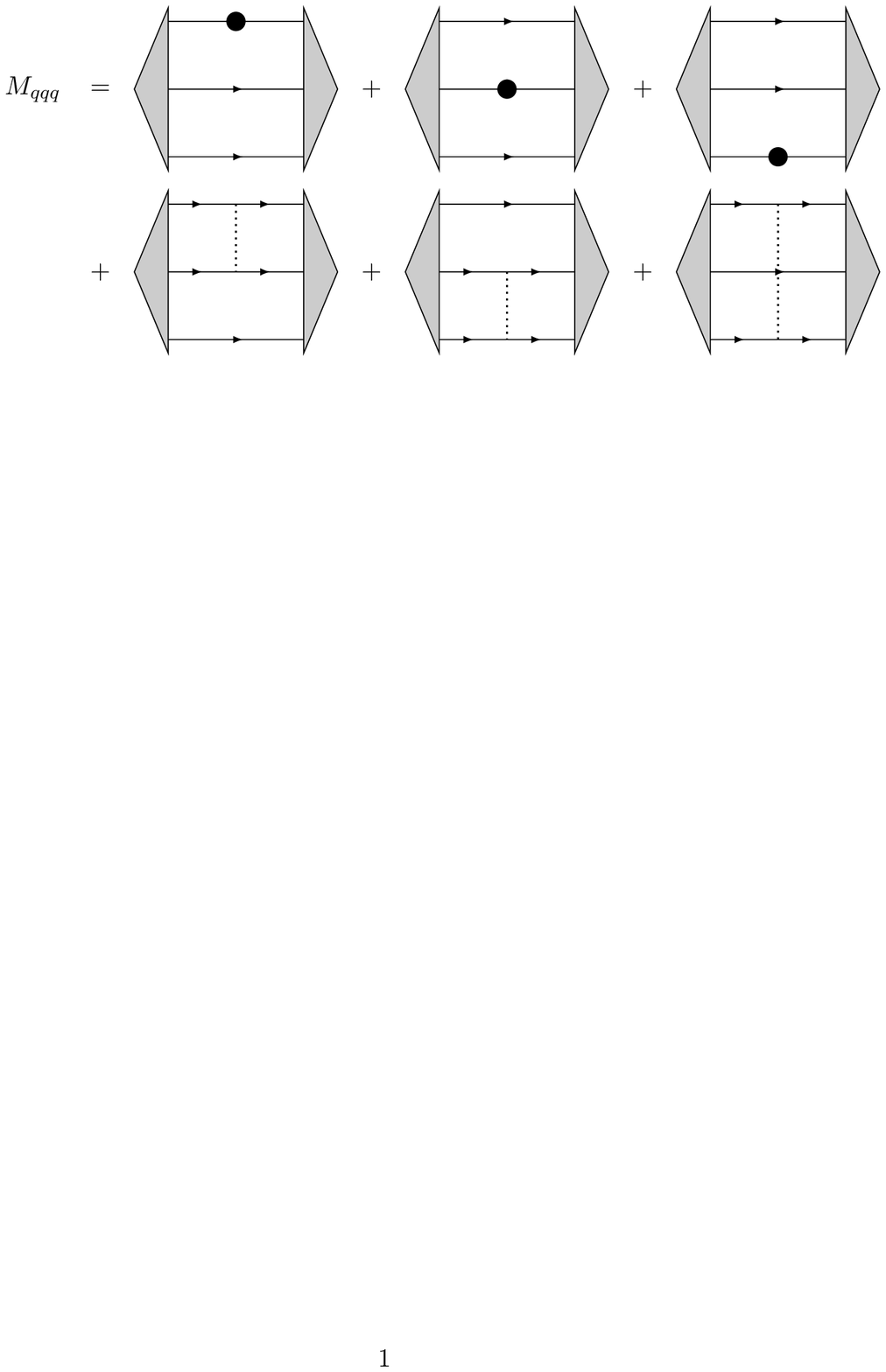}
\caption{Baryon diagrams for  $ \langle\psi_{qqq}|H_{phen}|\psi_{qqq}\rangle  $}\label{pic:diagram01}
\end{figure}

The proton is taken as an example for calculating  the baryon octet mass. According to the discussion above, the proton state can be expressed as
\be
|\text{proton}, \frac{1}{2}  \rangle
=\sum_{\mathcal C_1 \mathcal C_2\mathcal C_3}\int
\frac{d\mathbf k_1}{(2\pi)^3}   \frac{d\mathbf k_2}{(2\pi)^3}f(\mathbf k_1,\mathbf k_2,\mathbf k_3)
\times \psi_{fs}  (\text{proton},\frac{1}{2})
\times \varepsilon_{\mathcal C_1\mathcal C_2\mathcal C_3}
 b^{\dagger}_{\mathcal C_1 \lambda_1 u}(\mathbf{k}_1)
 b^{\dagger}_{\mathcal C_2 \lambda_2 u}(\mathbf{k}_2)
 b^{\dagger}_{\mathcal C_3 \lambda_3 d}(\mathbf{k}_3)
\arrowvert
0 \rangle,
\ee
where
\ba
\psi_{fs} (\text{proton},\frac{1}{2})&=&\frac{1}{3\sqrt 2} (\uparrow \downarrow \uparrow- \downarrow \uparrow \uparrow)(udu-duu)
+\frac{1}{3\sqrt 2} (\uparrow \uparrow \downarrow- \uparrow \downarrow \uparrow)(uud-udu)
+\frac{1}{3\sqrt 2} (\uparrow \uparrow \downarrow- \downarrow\uparrow  \uparrow)(uud-duu)   \notag \\
&=&\frac{2}{3\sqrt 2} u(\uparrow) u(\uparrow) d(\downarrow) -\frac{1}{3\sqrt 2} u(\uparrow) u(\downarrow) d(\uparrow) -\frac{1}{3\sqrt 2} u(\downarrow) u(\uparrow) d(\uparrow)   \notag \\
&+&\frac{2}{3\sqrt 2} u(\uparrow) d(\downarrow) u(\uparrow) -\frac{1}{3\sqrt 2} u(\downarrow) d(\uparrow) u(\uparrow) -\frac{1}{3\sqrt 2} u(\uparrow) d(\uparrow) u(\downarrow)     \notag \\
&+&\frac{2}{3\sqrt 2} d(\downarrow) u(\uparrow) u(\uparrow) -\frac{1}{3\sqrt 2} d(\uparrow) u(\downarrow) u(\uparrow) -\frac{1}{3\sqrt 2} d(\uparrow) u(\uparrow) u(\downarrow).
\ea

Contributions to the Hamiltonian expectation value are summrized in Fig.\ 1.
The expectation value for the proton mass is
\ba
\langle\text{proton}, \frac{1}{2} |H_{t} | \text{proton}, \frac{1}{2}  \rangle
=M_{kine}+M_{12}+M_{23}+M_{31},
\ea
where the kinetic energy has the form
\be
M_{kine}=\int \frac{d\mathbf k_1}{(2\pi)^3}   \frac{d\mathbf k_2}{(2\pi)^3}
(\sqrt{m^2_u+{\mathbf k_1}^2}+\sqrt{m^2_u+{\mathbf k_2}^2} +\sqrt{m^2_d+{\mathbf k_3}^2})
f^{2}(\mathbf k_1,\mathbf k_2,\mathbf k_3).
\ee
The matrix elements $M_{12}$, $M_{23}$, $M_{31}$ have many terms due to the complexity of proton flavor-spin wave function $\psi_{fs}$.
For example  $M_{12}$ is
\be
M_{12}=-\frac{2}{3}\int \frac{d\mathbf k_1}{(2\pi)^3}   \frac{d\mathbf k_2}{(2\pi)^3}   \frac{d\mathbf{q}}{(2\pi)^3}
f(\mathbf k_1,\mathbf k_2,\mathbf k_3)    f(\mathbf k^\prime_1,\mathbf k^\prime_2,\mathbf k^\prime_3)
V(|\mathbf q|)
(\frac{4}{18} E_1+\frac{2}{18} E_2+\frac{2}{18} E_3+\frac{8}{18} E_4+\frac{2}{18} E_5+\frac{2}{18} E_6-\frac{4}{18} E_7-\frac{4}{18} E_8).
\ee
The eight contributions $E_i$ are given in Table \ref{tab:elements}. They are classified by different spin configurations. The expressions for  $M_{13}$ ($M_{23}$) are similar to $M_{12}$ with the replacement of $k_2$ and $k^{'}_2$ by $k_3 $ and $k^{'}_3$ ($k_1$ and $k^{'}_1 $ by $k_3$ and $k^{'}_3$).
Due to the symmetry of the wave function, the numerical values of $M_{12}$, $M_{13}$ and $M_{23}$
are the same.

\begin{table}[t]
\centering
\caption{\label{tab:elements} Expressions  $E_i$ and  coefficients  $M_{12}$ for the proton.}
\begin{tabular}{c|c|c|c}
\hline
\hline
spin    &   contributing terms   &     coefficient      &       matrix element         \\
\hline
$\uparrow \uparrow \rightarrow \uparrow \uparrow$
 &  $u(\uparrow)u(\uparrow) \rightarrow u(\uparrow)u(\uparrow)$
 &  $\frac{4}{18} $
 &  $E_1=[  \bar u_{\frac{1}{2}} ({ {\bf k}_1}^{\prime}) \Gamma u_{\frac{1}{2}}({{\bf k}_1})  ]
[  \bar u_{\frac{1}{2}} ({ {\bf k}_2}^{\prime}) \Gamma u_{\frac{1}{2}}({{\bf k}_2})  ]$    \\

 &  $u(\uparrow)d(\uparrow) \rightarrow u(\uparrow)d(\uparrow)$
 &  $\frac{2}{18} $
 &  $E_2=[  \bar u_{\frac{1}{2}} ({ {\bf k}_1}^{\prime}) \Gamma u_{\frac{1}{2}}({{\bf k}_1})  ]
[  \bar d_{\frac{1}{2}} ({ {\bf k}_2}^{\prime}) \Gamma d_{\frac{1}{2}}({{\bf k}_2})  ]$    \\

\hline

$\uparrow \downarrow \rightarrow \uparrow \downarrow$
 &  $u(\uparrow)u(\downarrow) \rightarrow u(\uparrow)u(\downarrow)$
 &  $\frac{2}{18} $
 &  $E_3=[  \bar u_{\frac{1}{2}} ({ {\bf k}_1}^{\prime}) \Gamma u_{\frac{1}{2}}({{\bf k}_1})  ]
[  \bar u_{-\frac{1}{2}} ({ {\bf k}_2}^{\prime}) \Gamma u_{-\frac{1}{2}}({{\bf k}_2})  ]$    \\

 &  $u(\uparrow)d(\downarrow) \rightarrow u(\uparrow)d(\downarrow)$
 &  $\frac{8}{18} $
 &  $E_4=[  \bar u_{\frac{1}{2}} ({ {\bf k}_1}^{\prime}) \Gamma u_{\frac{1}{2}}({{\bf k}_1})  ]
[  \bar d_{-\frac{1}{2}} ({ {\bf k}_2}^{\prime}) \Gamma d_{-\frac{1}{2}}({{\bf k}_2})  ]$    \\

 &  $d(\uparrow)u(\downarrow) \rightarrow d(\uparrow)u(\downarrow)$
 &  $\frac{2}{18} $
 &  $E_5=[  \bar d_{\frac{1}{2}} ({ {\bf k}_1}^{\prime}) \Gamma d_{\frac{1}{2}}({{\bf k}_1})  ]
[  \bar u_{-\frac{1}{2}} ({ {\bf k}_2}^{\prime}) \Gamma u_{-\frac{1}{2}}({{\bf k}_2})  ]$    \\

\hline
$\uparrow \downarrow \rightarrow \downarrow \uparrow$
 &  $u(\uparrow)u(\downarrow) \rightarrow u(\downarrow)u(\uparrow)$
 &  $\frac{2}{18} $
 &  $E_6=[  \bar u_{-\frac{1}{2}} ({ {\bf k}_1}^{\prime}) \Gamma u_{\frac{1}{2}}({{\bf k}_1})  ]
[  \bar u_{\frac{1}{2}} ({ {\bf k}_2}^{\prime}) \Gamma u_{-\frac{1}{2}}({{\bf k}_2})  ]$    \\

 &  $u(\uparrow)d(\downarrow) \rightarrow u(\downarrow)d(\uparrow)$
 &  $-\frac{4}{18} $
 &  $E_7=[  \bar u_{-\frac{1}{2}} ({ {\bf k}_1}^{\prime}) \Gamma u_{\frac{1}{2}}({{\bf k}_1})  ]
[  \bar d_{\frac{1}{2}} ({ {\bf k}_2}^{\prime}) \Gamma d_{-\frac{1}{2}}({{\bf k}_2})  ]$    \\

 &  $d(\uparrow)u(\downarrow) \rightarrow d(\downarrow)u(\uparrow)$
 &  $-\frac{4}{18} $
 &  $E_8=[  \bar d_{-\frac{1}{2}} ({ {\bf k}_1}^{\prime}) \Gamma d_{\frac{1}{2}}({{\bf k}_1})  ]
[  \bar u_{\frac{1}{2}} ({ {\bf k}_2}^{\prime}) \Gamma u_{-\frac{1}{2}}({{\bf k}_2})  ]$    \\

\hline
\hline

\end{tabular}
\end{table}

\begin{table}[t]
\centering
\caption{\label{tab:elements2}The expressions  $E^{'}_1$, $E^{'}_2$ and  coefficients  $M^{'}_{12}$ for the $\Delta^{+}$.}
\begin{tabular}{c|c|c|c}
\hline
\hline
spin    &   contributing terms   &     coefficient      &       matrix element         \\
\hline
$\uparrow \uparrow \rightarrow \uparrow \uparrow$
 &  $u(\uparrow)u(\uparrow) \rightarrow u(\uparrow)u(\uparrow)$
 &  $\frac{1}{3} $
 &  $E^{'}_1=[  \bar u_{\frac{1}{2}} ({ {\bf k}_1}^{\prime}) \Gamma u_{\frac{1}{2}}({{\bf k}_1})  ]
[  \bar u_{\frac{1}{2}} ({ {\bf k}_2}^{\prime}) \Gamma u_{\frac{1}{2}}({{\bf k}_2})  ]$    \\

 &  $u(\uparrow)d(\uparrow) \rightarrow u(\uparrow)d(\uparrow)$
 &  $\frac{2}{3} $
 &  $E^{'}_2=[  \bar u_{\frac{1}{2}} ({ {\bf k}_1}^{\prime}) \Gamma u_{\frac{1}{2}}({{\bf k}_1})  ]
[  \bar d_{\frac{1}{2}} ({ {\bf k}_2}^{\prime}) \Gamma d_{\frac{1}{2}}({{\bf k}_2})  ]$    \\

\hline
\hline
\end{tabular}
\end{table}


For the decuplet states the  $\Delta^{+}$ is used as a representative example and has wavefunction
 given by
\be
|\Delta^{+}, \frac{3}{2} \rangle
=\sum_{\mathcal C_1 \mathcal C_2\mathcal C_3}\int
\frac{d\mathbf k_1}{(2\pi)^3}   \frac{d\mathbf k_2}{(2\pi)^3}f(\mathbf k_1,\mathbf k_2,\mathbf k_3)
\times \psi_{fs}(\Delta^{+},\frac{3}{2})
\times \varepsilon_{\mathcal C_1\mathcal C_2\mathcal C_3}
 b^{\dagger}_{\mathcal C_1 \lambda_1 u}(\mathbf{k}_1)
 b^{\dagger}_{\mathcal C_2 \lambda_2 u}(\mathbf{k}_2)
 b^{\dagger}_{\mathcal C_3 \lambda_3 d}(\mathbf{k}_3)
\arrowvert
0 \rangle,
\ee
in which
\ba
\psi_{fs}(\Delta^{+},\frac{3}{2})=\frac{1}{\sqrt 3}[u(\uparrow)u(\uparrow)d(\uparrow)+u(\uparrow)d(\uparrow)u(\uparrow)
+d(\uparrow)u(\uparrow)u(\uparrow)],
\ea
and the mass is given by
\ba
\langle\Delta^{+}, \frac{3}{2}|H_{t} | \Delta^{+}, \frac{3}{2}\rangle
=M^{'}_{kine}+M^{'}_{12}+M^{'}_{23}+M^{'}_{31},
\ea
where
\be
M^{'}_{kine}=\int \frac{d\mathbf k_1}{(2\pi)^3}   \frac{d\mathbf k_2}{(2\pi)^3}
(\sqrt{m^2_1+{\mathbf k_1}^2}+\sqrt{m^2_2+{\mathbf k_2}^2} +\sqrt{m^2_3+{\mathbf k_3}^2})
f^{2}(\mathbf k_1,\mathbf k_2,\mathbf k_3),
\ee
\be
M^{'}_{12}=-\frac{2}{3}\int \frac{d\mathbf{k_1}}{(2\pi)^3}   \frac{d\mathbf{k_2}}{(2\pi)^3}   \frac{d\mathbf{q}}{(2\pi)^3}
f(\mathbf k_1,\mathbf k_2,\mathbf k_3)    f(\mathbf k^\prime_1,\mathbf k^\prime_2,\mathbf k^\prime_3)
V(|\mathbf q|)
[E^{'}_1+E^{'}_2],
\ee
The expressions for $E^{'}_1$ and $E^{'}_2$ are given in Table II.
Again, due to the symmetry of the wave function, the numerical values for $M^{'}_{13}$ and $M^{'}_{23}$ are the same as $M^{'}_{12}$ .

\begin{table}[t]
\centering
\caption{Unified model parametersl. The different meson \cite{Godfrey:1985xj} and baryon \cite{Capstick:1986bm}  parameters  from Isgur et al.  are
also listed.}
\begin{tabular}{c c c c c}
\hline
\hline
Parameters & \quad $\xi=2.1$  & \quad $\xi=1.0$  &  \quad Ref.\ \cite{Godfrey:1985xj} & \quad Ref.\ \cite{Capstick:1986bm} \\
\hline
$m_u/m_d$ (MeV)  & 313 & 50 & 220 &220 \\
$m_s$  (MeV)            & 660  &  640 & 419  & 419 \\
$\alpha_s$     & 0.40 & 0.40   & 0.60  & 0.60 \\
$\sigma$ (GeV$^2$)    & 0.18  & 0.18 &0.18  & 0.15 \\
$C$ (MeV)                     &-195  &-198 &-253 &-615  \\
\hline
$\alpha_1$                  & 0.762 & 0.490    &       &     \\
$\sigma_1$ (GeV$^2$)    & 0.207 & 0.0625  &      &     \\
\hline
\hline
\end{tabular}
\end{table}

\section {Numerical results}\label{sec:results}

The new spin splitting interaction Hamiltonian $H_{I1}$
was investigated by calculating
the meson and baryon masses  for all possible $\Gamma$ matrices.
The interaction with matrices $1$ and $\gamma_5$ invert the baryon octet and decuplet spectra, i.e. produce larger octet masses than decuplet masses. The matrices $\vec{\gamma}\gamma_0,\gamma_5\gamma_0,$ and $\vec{\gamma}\gamma_5$ produce $P$ wave meson masses several hundred MeV lower than the experimental values.
Only the interaction with the  $\Gamma=\vec{\gamma}$ could produce reasonable baryon  and meson masses simultaneously. This is the same Lorenz structure as in Ref.\ \cite{LlanesEstrada:2004} using an effective one gluon exchange hyperfine interaction.
Then using the predetermined values  $\sigma = .18$ GeV$^2$ and $\alpha_s = .4$, the remaining free model parameters, the
 $u/d, s$ quark masses, the potential strengths $\sigma_1$ and $\alpha_1$ in $H_{I1}$ and constant $C$ in $H_{I0}$, were determined by reproducing the light quark meson spectrum.
These parameters are listed in Table III along  with Godfrey and Isgur's quark  model values for comparison.
The inclusion of the meson-exchange interaction does not add a free parameter.

In this model the  relativistic four-component  spinor $u_\lambda$ is related to the free quark
propagator by
\begin{equation}
\sum_{\lambda=1,2} u_{\lambda}({\bf k})  \bar{u}_{\lambda}({\bf k}) = \slash{p} + m.
\end{equation}
However for confined quarks  lattice results obtain a different propagator so modified spinors  of the form
\begin{eqnarray}
 u_{\lambda}({\bf k})&=&\frac{1}{\sqrt 2}
 \left(
  \begin{array}{c}
    \sqrt{\frac{\omega+\xi m}{\omega}}\chi_{\lambda}\\
     \sqrt{\frac{\omega-\xi m}{\omega}}\boldsymbol{\sigma}\cdot\hat{\bf k}\chi_{\lambda}
  \end{array}
  \right),\\
 v_{\lambda}({\bf k})&=&\frac{1}{\sqrt 2}
 \left(
  \begin{array}{c}
    \sqrt{\frac{\omega-\xi m}{\omega}} \boldsymbol{\sigma}\cdot\hat{\bf k} \chi_{\lambda}\\
     \sqrt{\frac{\omega+\xi m}{\omega}}\chi_{\lambda}
  \end{array}
  \right),
\end{eqnarray}
have also been investigated. Here
 $\omega=\sqrt{k^2 + \xi^2 m^2}$ with parameter $\xi$. The free quark spinor is obtained for $\xi=1$
while  $\xi \rightarrow \infty$ produces  the non-relativistic two component spinor $\chi_\lambda$.
\begin{table}[t]
\centering
\caption{Calculated hadron spectrum  in MeV. The experimental values from PDG are listed
in the last column.}\label{table:data}
\begin{tabular}{ccccc}
\hline
\hline
$J^{PC}$   &\ \quad Meson     &\ \quad This work($\xi=2.1$)  & \ \ \quad   This work($\xi=1.0$)   & \ \quad PDG  \\
\hline
               & $\pi$   &  141       & 137   & 135           \\

 $0^{-+}$   & $K$   &  494     & 498       &  493        \\

\hline
              & $\rho$   &  778 & 779  &776  \\

 $1^{--} $    & $K^*$   &  891 & 888& 894   \\

            & $\phi$   &  1029 & 995 &1020  \\
\hline
            & $b_1$   & 1195 & 1043 & 1235   \\

 $1^{+-}$  &$K_{1B}$   &  1346 &  1277    &  \\

            & $h_1$   &  1512 & 1485 &1380    \\
\hline
            & $a_0$   &  1460 & 1352 &1450     \\

 $0^{++}$  &$K_0^*$   & 1519 & 1473 & 1430  \\

            & $f_0$   &  1623 &1609 &1710   \\
\hline
 $\frac{1}{2}^{+}$               & $N$   &  934      & 941 & 938                 \\

                & $\Lambda$   &  1158    &  1180       & 1116        \\

                & $\Sigma$   &  1204     & 1196        & 1189        \\

               & $\Xi$   & 1360 &    1340    & 1314         \\
\hline
 $\frac{3}{2}^{+}$               & $\Delta$   &  1233      & 1254 & 1232                 \\

                & $\Sigma^*$   &  1385     & 1400        & 1385        \\

               & $\Xi^*$   & 1544 &    1543   &1533           \\

               & $\Omega$   & 1704 &    1681    &1672        \\
\hline
\hline
\end{tabular}
\end{table}
The hadron spectra were studied for different values of $\xi$ and quark masses and the results are listed in Tables III and IV. While different sets of values produce comparable hadron spectra, the value $\xi = 2.1$ yields quark masses similar to the constiuent quark model while for $\xi = 1$ a much lighter $u$ quark mass
of 50 MeV is required. Note the quark masses are purely parameters which should not necessarily be identified as constituent quark masses.

The hadron masses are obtained by variationally using using the Monte Carlo method and are compared to experiment in Table \ref{table:data}.
 Only meson states suggested as $q\bar{q}$  in the PDG review table (Table 14.2) \cite{PDG} are addressed.
For mesons the Goldstone exchange interaction is small, less than 20 MeV.
However for baryons it is larger,  reducing the decuplet masses by about 30 MeV and for octets  between
60 and 100 MeV which is now sufficient to reproduce the observed 300 MeV $N\Delta$ mass splitting. This is gratifying because this splitting  without Goldstone exchange is only 250 MeV. It appears Goldstone exchange interactions play an important role in the baryon spectrum \cite{Theberge}.

The calculated meson masses agree quite well with PDG data, especially the $0^{-+}$  and  $1^{--}$ states.
The $\pi  \rho$ splitting is close to 640 MeV and the $K K^*$ splitting is about 400 MeV. In traditional quark models these
splittings are produced by the color hyperfine interaction. In this work it is predominantly obtained from the Hamiltonian $H_{I1}$ which lowers the $0^{-+}$ masses by about 500 MeV while reducing  $1^{--}$ masses  less than 100 MeV.

The model parameters were mainly determined by the $\pi$, $\rho/\omega$, $K$ and $K^*$ masses and then the remaining hadron masses were  predicted.
With the exception of three states ($h_1$, $K^*_0$ and $f_0$) the overall meson and baryon specta are in very good agreement with observation.
This model calculation also predicts that the lightest scalar mesons have mass well above 1 GeV.  This would indicate that the $a_0(980)$, $f_0(980)$
and $f_0(500)$ mesons are not pure $q\bar{q}$ states and  possibly have a tetraquark structure.

\section{Summary}\label{sec:conclusion}

A new, unified Hamiltonian model has been developed which combines the attractive features of
phenomenologically based  quark models with many of the theoretical ingredients common to QCD.
A new spin interaction has also been investigated for a variety of Lorentz structures with a clear
preference for $\Gamma = \vec{\gamma}$. A Goldstone exchange interaction was also included and,
along with the spin interaction, found necessary to
accurately reproduce the $N\Delta$ mass splitting.

The parameters are mainly determined by fitting the $\pi$, $\rho/\omega$, $K$ and $K^*$ masses.
The remaining meson and baryon masses were then predicted and found to be in good agrrement with observation.
All scalar mesons are predicted to have mass well above 1 GeV suggesting the
$a_0(980)$, $f_0(980)$ and $f_0(500)$ are not simple  $q\bar{q}$ states but perhaps tetraquarks.
Most significantly, a good Hamiltonian description for the meson,  baryon (octet and decuplet) spectra has been 
obtained with a common set of  parameters which has previously not been achieved.

Future work will address heavy quark systems to further test this model.  If robust results are obtained
applications to exotic systems will be performed.

\section*{Acknowledgments}

The authors are grateful to S. R. Cotanch for helpful discussions.
This work is supported in part by DFG and NSFC (CRC 110) and by the
National Natural Science Foundation of China (Grant No. 11035006).

\end{document}